# Influence of stress on magneto-impedance in $Co_{71-x}Fe_xCr_7Si_8B_{14}$ (x = 0, 2) amorphous ribbons

B. Kaviraj and S. K. Ghatak[1]

Department of Physics & Meteorology, Indian Institute of Technology, Kharagpur 721302, India

[1] **Corresponding Author**
**skghatak@phy.iitkgp.ernet.in**

Abstract

Systematic measurements of stress-impedance (SI) have been carried out using Co-rich amorphous ribbons of nominal composition $Co_{71-x}Fe_xCr_7Si_8B_{14}$ (x = 0, 2) at various excitation frequencies and bias fields and at room temperature. The impedance, Z(σ) for both the samples was found to be very sensitive functions of applied tensile stresses (up to 100MPa) exhibiting a maximum SI ratio as much as 80% at low frequency ~ 0.1MHz. The nature of variation of SI changes with the excitation frequency especially at higher frequencies in MHz region where it exhibits a peak. Magnetization measurements were also performed to observe the effects of applied stresses and magnetization decreased with the application of stress confirming the negative magnetostriction co-efficient of both the samples. Both the samples exhibited negative magneto-impedance (MI) when the variation of Z is observed with the applied bias magnetic field, H. The impedance as functions of applied magnetic field, Z(H), decreases with the application of stress thus making the MI curves broader. Maximum MI ratio as large as 99% has been observed for both the samples at low fields ~ 27Oe.

## I. INTRODUCTION

The magneto-impedance (MI) effect – a large change of impedance induced by d.c. magnetic field is normally observed in a soft ferromagnetic conductor and in most situations the impedance is reduced to a large extent compared to that in a zero field state. The ferromagnetic materials like the amorphous transition metal-metalloid alloys, with the shapes of wires [1-3], ribbons [4-7] and thin films [8-9], are forerunners in exhibiting large negative MI at low frequency [10-13]. The Fe-or Co-based metallic glasses in the form of ribbons or wires are magnetically soft and the relative decrease in impedance in presence of small fields is very large [10-14]. The MI effect has been observed for as-quenched, amorphous as well as in the nanocrystallized materials [15-17]. The MI effect is associated with the field penetration of electromagnetic (e.m) waves within a magnetic metal with high permeability. The impedance of a metal is determined by penetration depth of e.m. field and the penetration depth in turn is a measure of the screening of field. In paramagnetic metals, the screening is solely due to the conduction electrons and an additional component of screening arises from the a.c. magnetization current in ferromagnetic metals. This leads to higher impedance in ferromagnetic state. However, screening of magnetic origin can be altered significantly by changing the magnetic response to a.c field. The response, as measured by the dynamic permeability $\mu$, can be altered by applying an external d.c. magnetic field, external stress, thermal treatment etc. In most cases, a moderate magnetic field increases penetration depth of the e.m. field at r.f range and thereby the impedance is decreased compared to that at zero field [18-19]. The permeability of a ferromagnetic substance is governed by the magnetic anisotropy energy. The external stress can modify the anisotropy in a material with non-zero magnetostriction coefficient and therefore induce a change in impedance – referred as stress-impedance [SI] [20-22]. The Co-rich ferromagnetic amorphous alloys have high permeability, low magnetic losses and low magnetostriction constant, and these make the materials suitable for observing both SI and MI effects [20, 21].

In this communication, the influence of stress on magneto-impedance and magnetization are studied for the sample $Co_{71-x}Fe_xCr_7Si_8B_{14}$ (x= 0, 2) at different excitation frequencies and bias fields.

## II. EXPERIMENTAL DETAILS

Amorphous ribbons with nominal compositions $Co_{71}Cr_7Si_8B_{14}$ and $Co_{69}Fe_2Cr_7Si_8B_{14}$ were produced using the melt-spinning technique. The sample cross-sections were 6.41x0.0335 and 6.347x0.0325 $mm^2$ respectively. All the samples were cut in 8 cm long pieces for the measurements. All the samples were used in as-quenched state and were aligned with their longitudinal axis perpendicular to the Earth's magnetic field. The samples were placed within a small signal coil of rectangular geometry and with 100 turns. The coil was located at the middle of the sample-length which was larger than the length of the coil in order to reduce the demagnetization effects. The coil was connected to the current terminal of the Impedance Analyzer (Model-HP4294A) where a sinusoidal current amplitude was kept constant at $I_{rms}$ = 20mA. This created an a.c magnetic field of about 196A/m along the axis of the coil which is also along the length of the ribbon. The voltage response of the sample around this field was found to be linear. The frequency of the driving current was scanned from 100KHz to 10MHz. Maximum d.c bias field up to

27Oe was applied parallel to the exciting a.c field and to the length of the ribbon. Magnetization measurements were performed with the help of an a.c magnetometer at 70Hz frequency.

These samples were further subjected to tensile stresses using load and pulley arrangement. Maximum stresses up to 100MPa were applied in the longitudinal direction (which is also the direction of exciting a.c field) depending upon the cross-section of the ribbon. The real and imaginary components of impedance Z = R+jX where X= ωL (L being the inductance of the sample) were measured across the signal coil with the help of Impedance Analyzer. The resistance and reactance of the empty coil and test leads were subtracted and only the change in impedance due to the sample was taken into account. All measurements were performed at room temperature. The stress-impedance ratio has been expressed as: $\delta Z_{\sigma}\% = \frac{Z(\sigma) - Z(0)}{Z(0)} \times 100$.

## III. RESULTS

### A. Sample $Co_{71}Cr_7Si_8B_{14}$ (x = 0)

Fig (1) shows the excitation frequency response of the resistive (R) and reactive (X) components of impedance for the sample $Co_{71}Cr_7Si_8B_{14}$ as functions of different external tensile stresses. At all stresses, R and X are low at low frequencies but increase monotonically at higher frequencies. The reactive part is greater than the resistive part. The applied tensile stress (σ) decreases the magnitude of R progressively as the stress is increased from 0MPa to 100MPa. The frequency response of the reactive component of Z on the other hand exhibits a 'cross-over' in high frequency region (> 1MHz) at σ =

20MPa where it crosses the 'zero-stress' curve indicating that the values of X at 20MPa stress are greater than those at 'zero-stress' in this frequency region. But at σ = 100MPa, the values of X are lower than those at σ = 0MPa. These will be clear from the results of stress impedance measurements that are presented subsequently.

In Fig(2), we show the stress impedance results for the sample $Co_{71}Cr_7Si_8B_{14}$ measured at different excitation frequencies of 0.1MHz, 1MHz and 10MHz. Here the percentage change in impedance $\delta Z_\sigma\%$ has been plotted as functions of tensile stresses. At low frequencies, the impedance is maximum at zero stress and falls off sharply with the increase of stress. At a higher frequency (10MHz), the impedance increases with stress and exhibits a maximum at σ ~ 20MPa and $\delta Z_\sigma\%$ becomes negative at higher stress. The maximum relative change in SI decreases from 80% (0.1MHz) to around 40% (10MHz) with the increase in frequency in the range of applied stresses.

The impedance (Z) has also been measured at different d.c biasing fields (H) at 1MHz frequency and the results for $\delta Z_H\% = \frac{(Z(H)-Z(0))}{Z(0)}\times 100$ versus H are depicted in Fig. 3 for σ = 0, 32, 65, 90MPa. The impedance of the sample exhibits a maximum at zero d.c field and decreases monotonically with H, thus exhibiting a negative magneto-impedance phenomenon. In the absence of stress (σ = 0), the impedance of the sample has been found to drop from 36Ω (H = 0) to 0.02Ω (H = 27Oe) with the application of field. The impedance saturates to very low values for H > 20Oe. The maximum change in magneto-impedance (MI) ratio with respect to zero-field was found to be nearly 99% for all values of applied stresses. With the increase of applied stress, the MI response becomes flatter and it saturates at higher values of applied d.c field. At higher frequencies, the observed field dependence of Z was found to remain the same but the maximum MI ratio decreased at higher values of stresses.

**B. Sample $Co_{69}Fe_2Cr_7Si_8B_{14}$ (x = 2)**

The frequency response of the resistive and reactive components of impedance for this particular ribbon is shown in Fig (4). Fig (4) shows the monotonic increase of resistive and reactive components of impedance at high frequencies. The 'cross-overs' with the $\sigma = 0$ curve suggests that within this stress interval, the relative change in SI is positive.

Fig (5) depicts the relative change in SI for the sample $Co_{69}Fe_2Cr_7Si_8B_{14}$ as functions of different stresses at different excitation frequencies (0.1, 1 and 10MHz). Comparing with the SI response of the previous sample in Fig (2), we notice that in this sample the maximum relative changes in SI are lower at all the chosen range of frequencies. The maximum change is around 70% and 47% at 0.1 and 1MHz respectively. At 1MHz and 10MHz, the SI% becomes positive (hence exhibits a peak) at finite intervals of stresses and thereafter decreases. The stress value at which SI% shows a maxima increases from $\sigma \sim 9$MPa at 1MHz to $\sigma \sim 30$MPa at 10MHz. Also compared to the previous sample, we note that in this sample, the SI% remains positive for a larger extent of stress at high frequencies ~ 10MHz. It is positive for the entire range of the chosen interval of stress.

In Fig. 6, the field dependence of magneto-impedance has been plotted as functions of different external stresses and at an excitation frequency of 1MHz. This sample also exhibits negative magneto-impedance for all values of applied stresses. Maximum MI ratio up to 99% has been observed for all values of stresses. We note that the zero-field impedance values for $\sigma = 32$MPa is larger than the corresponding value for $\sigma = 0$MPa, thus the sample exhibiting positive SI ratio within this interval of applied stress (see inset). The zero-field impedance values for higher stresses ($\sigma$ >32MPa) are lower than that of zero-stress. This also supports the observations in Fig. 5 where the SI ratio for 1MHz frequency remained positive up to $\sigma$ ~30MPa and becomes negative thereafter with the increase of stress.

The magnetization measurements of both the samples have been performed by fluxmetric method and at a frequency of 70Hz. Different external stresses also have been applied to study the effects of the change in the magnetization curves. They are depicted in the figures below (Fig. 7(a) and Fig. 7(b)). Figs. 7(a) and 7(b) show that the magnetizations for both the samples decrease with the application of external stress.

## IV. Discussions

From Figs. (2) and (5), we note that the impedance decreases with the application of stress. This decrease is monotonic at frequencies up to 1MHz for the sample with x = 0 and even lower up to KHz region for the sample with x = 2. This is because the Co-rich samples possess very small (near to zero) and negative magnetostriction coefficient. Increasing stress (tensile) in them promotes the growth of domains in the perpendicular direction, which in our geometry is the direction transverse to the long axis of the ribbon (and also transverse to the exciting a.c field). This causes a reduction in the longitudinal permeability (as the exciting field is along the length of the ribbon) and hence a reduction of impedance of the ribbon. Moreover at higher excitation frequencies, the sharpness of SI curves decreases. This is attributed to the decrease of permeability at high frequencies.

The magnetizations for both the samples decrease with the application of external stress (see Figs. 7(a) and (b)) because of the negative magnetostriction coefficient of the samples. The calculated values of saturation magnetization ($M_s$) were $1.77 \times 10^5$A/m and $2.47 \times 10^5$A/m for x = 0 and x = 2 ribbons respectively. Such high values of $M_s$ are close to those that are reported for Co-rich amorphous ribbons [22-23]. Comparing the two samples, we notice that the magnitudes of magnetization are higher for the sample $Co_{69}Fe_2Cr_7Si_8B_{14}$ (Fig. 7(b)) than the one having composition $Co_{71}Cr_7Si_8B_{14}$ (Fig. 7(a)). This is because of the fact that the magnetostriction coefficient ($\lambda_s$) of the pure Cobalt sample (x = 0) is higher than that of $Co_{69}Fe_2Cr_7Si_8B_{14}$ (x = 2). The estimated $\lambda_s$ from the slope of anisotropy energy ($E_k$) versus applied stress ($\sigma$) curves was found to be $-0.63 \times 10^{-6}$ and $-0.24 \times 10^{-6}$ for x = 0 and 2 respectively. This is shown in Fig. 8. These values were close to the measured values of $\lambda_s$ for Co-rich alloys [24, 25]. (Subtle variations in the measured values of $\lambda_s$ may arise from the internal stresses that are produced during the quenching process of the amorphous material). Since the anisotropy of amorphous magnetic materials is predominantly magnetostrictive in nature, the pure Cobalt sample possesses higher anisotropy energy and hence lower values of susceptibility than that of $Co_{69}Fe_2Cr_7Si_8B_{14}$.

The comparison of the SI response at different frequencies (100KHz and 10MHz) of the two samples is shown in Fig. 9. The results show that the SI response of the sample with x = 2 is broader compared to the sample with x = 0 in both the frequency regimes. The difference in broadness of the SI curves increases as we go to the higher frequency. The results can be explained by the fact that the magnetostriction coefficients of both the samples are negative but lower in magnitude and close to zero in the sample with x = 2.

The anisotropy energy in amorphous ribbons arises mainly from the coupling of the internal stresses with the magnetization due to the magnetoelastic effect and is lower in the sample x = 2 compared to that in x = 0. A transverse component of magnetic anisotropy inadvertently exists in the as-cast sample [9]. Since in a negative magnetostrictive sample, the application of stress promotes the growth of domains perpendicular to the direction of stress, the rate of decrease of permeability (longitudinal) must be greater for the sample with higher negative magnetostriction. Hence the fall of impedance with stress is also sharper for the sample with higher negative magnetostriction giving rise to sharper SI curves.

## V. CONCLUSIONS

Systematic measurements of stress-impedance and magneto-impedance for the samples $Co_{71-x}Fe_xCr_7Si_8B_{14}$ (x = 0, 2) lead us to conclude

(i) the variation of impedance with stress is a function of excitation frequency. At low frequency ~ 100KHz, impedance decreases monotonically with the applied stress but at high frequencies at or above 1MHz, the impedance rises to a maximum and then decreases at higher stress.

(ii) the SI curves are sharper for the sample with a higher magnitude of negative magnetostriction coefficient (x = 0). This is due to the larger decrease of longitudinal permeability with stress compared to that in the sample having lower magnitude of negative magnetostriction coefficient (x = 2).

(iii) due to the presence of very small and negative magnetostriction in both the alloys, very large and sharp changes in magneto-impedance has been observed with the application of very low biasing d.c fields, H ~ 20Oe. Maximum changes in magneto-impedance as large as 99% has been observed in both the alloys and the sharpness of magneto-impedance response decreases with the increase of applied stress.

## ACKNOWLEDGEMENTS

The authors are grateful to Dr. A. Mitra for providing the samples and S. Ghosh for technical help.

## List of Figure Captions

Fig 1. Frequency response of resistive (a) and reactive (b) components of impedance as functions of different tensile stresses for the sample $Co_{71}Cr_7Si_8B_{14}$ (x = 0).

Fig. 2. The variation of relative change in impedance with stress denoted by $\delta Z_\sigma \%$ for the sample $Co_{71}Cr_7Si_8B_{14}$ (x = 0) at 0.1, 1 and 10MHz frequency.

Fig.3.The relative change in magneto-impedance (%) denoted by $\delta Z_H \% = \frac{(Z(H)-Z(0))}{Z(0)} \times 100$ as functions of different external stresses for the sample $Co_{71}Cr_7Si_8B_{14}$ (x = 0) at 1MHz frequency. Inset shows the field dependence of absolute values of impedance.

Fig 4. Frequency response of resistive (a) and reactive (b) components of impedance as functions of different tensile stresses for the sample $Co_{69}Fe_2Cr_7Si_8B_{14}$ (x = 2).

Fig.5. The variation of relative change in impedance $\delta Z_\sigma \% = \frac{Z(\sigma)-Z(0)}{Z(0)} \times 100$ with stress for the sample $Co_{69}Fe_2Cr_7Si_8B_{14}$ (x = 2) at 0.1, 1 and 10MHz frequency.

Fig.6. The relative change in magneto-impedance (%) denoted by $\delta Z_H \% = \frac{(Z(H)-Z(0))}{Z(0)} \times 100$ as functions of different external stresses for the sample $Co_{69}Fe_2Cr_7Si_8B_{14}$ (x = 2) at 1MHz frequency. Inset shows the field dependence of absolute values of impedance.

Fig.7. Magnetization Curves for the sample $Co_{71}Cr_7Si_8B_{14}$ (a) and $Co_{69}Fe_2Cr_7Si_8B_{14}$ (b) as functions of different tensile stresses.

Fig. 8. The variation of the effective anisotropy, $E_k = \int_0^{M_s} H dM$ with stress σ for $Co_{71}Cr_7Si_8B_{14}$ (a) and $Co_{69}Fe_2Cr_7Si_8B_{14}$ (b) alloy.

Fig. 9. Comparison of SI response of the samples (x = 0, 2) at excitation frequencies of 100KHz (a) and 10MHz (b).

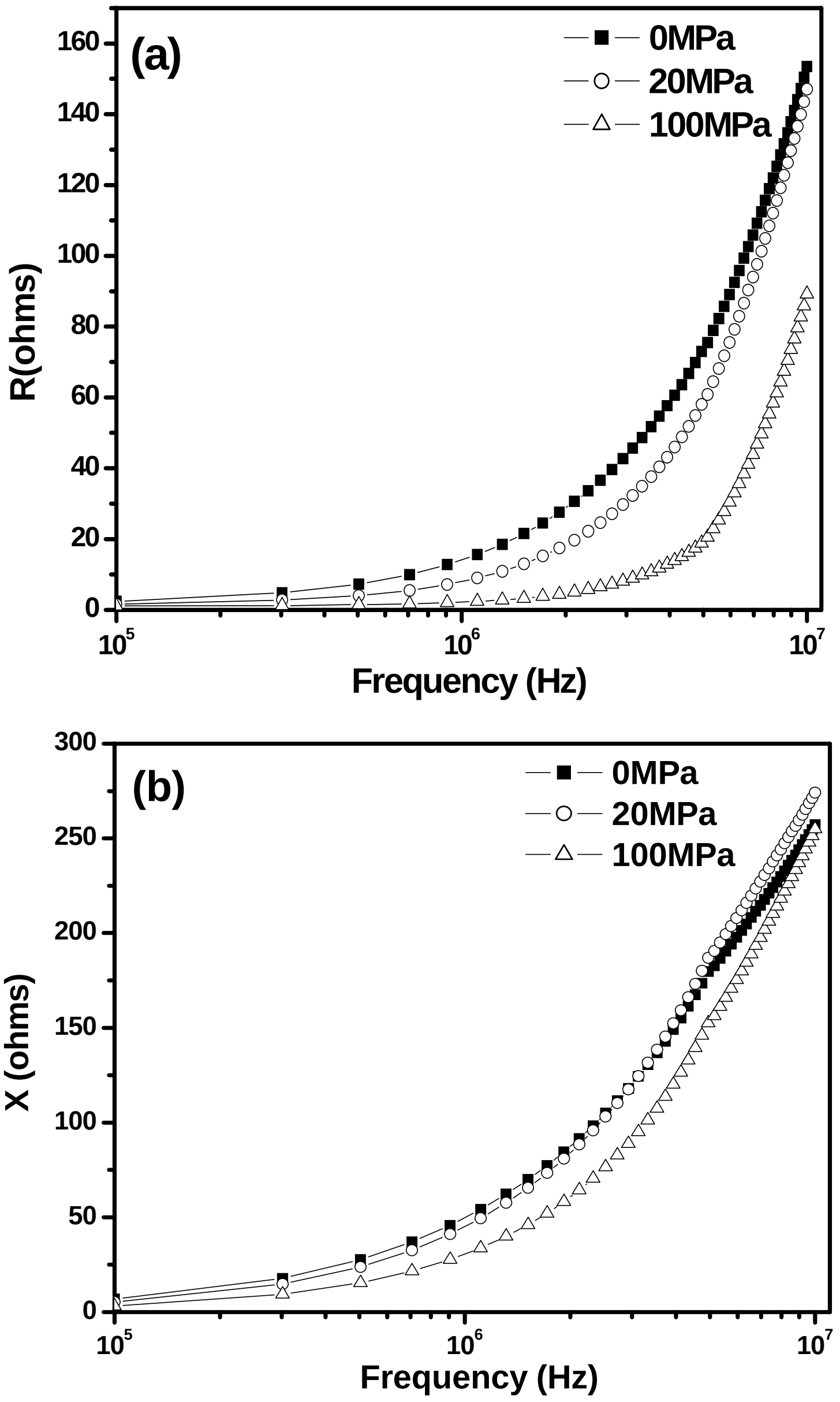

**Fig. 1. B. Kaviraj et al.**

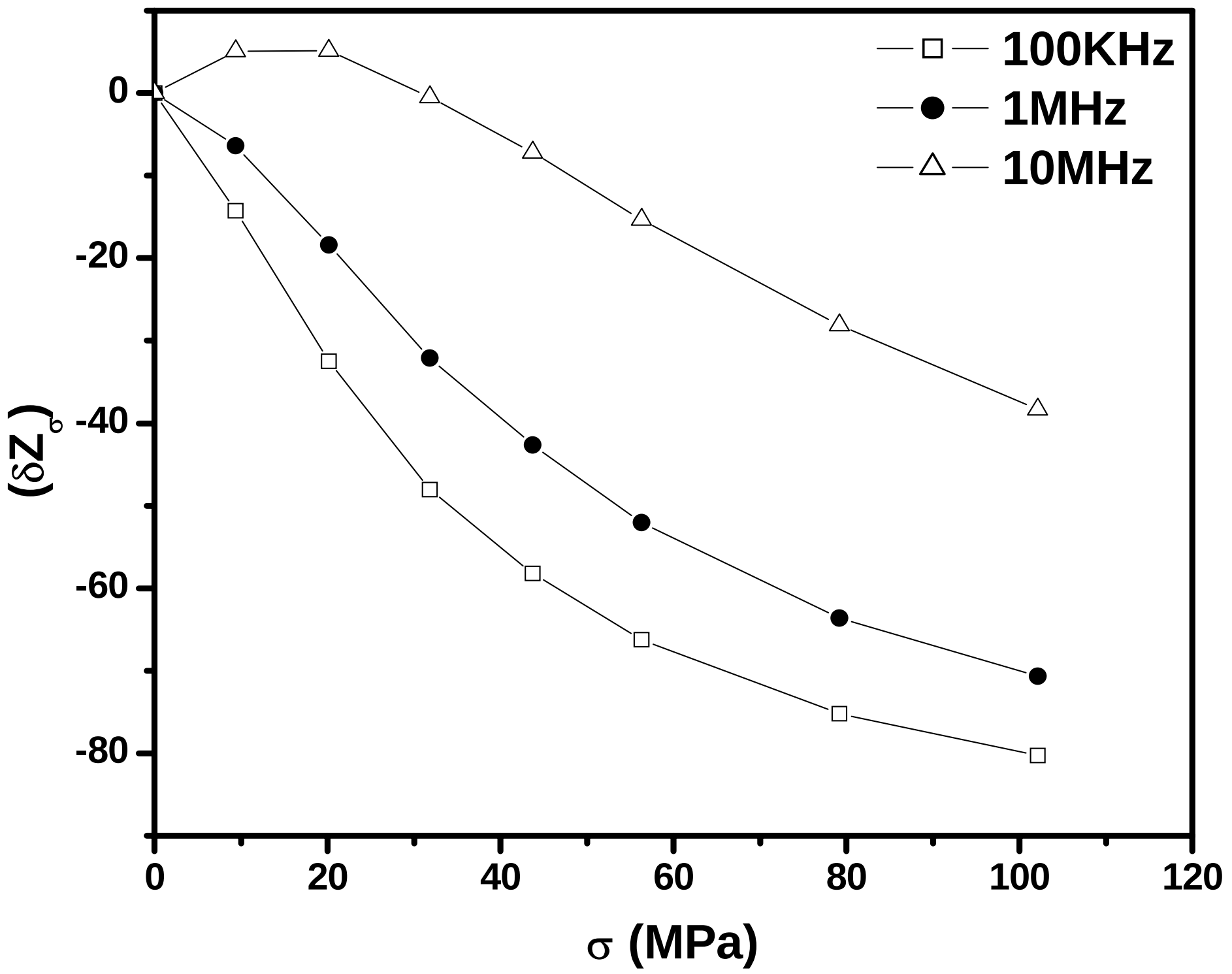

Fig. 2. B. Kaviraj et al.

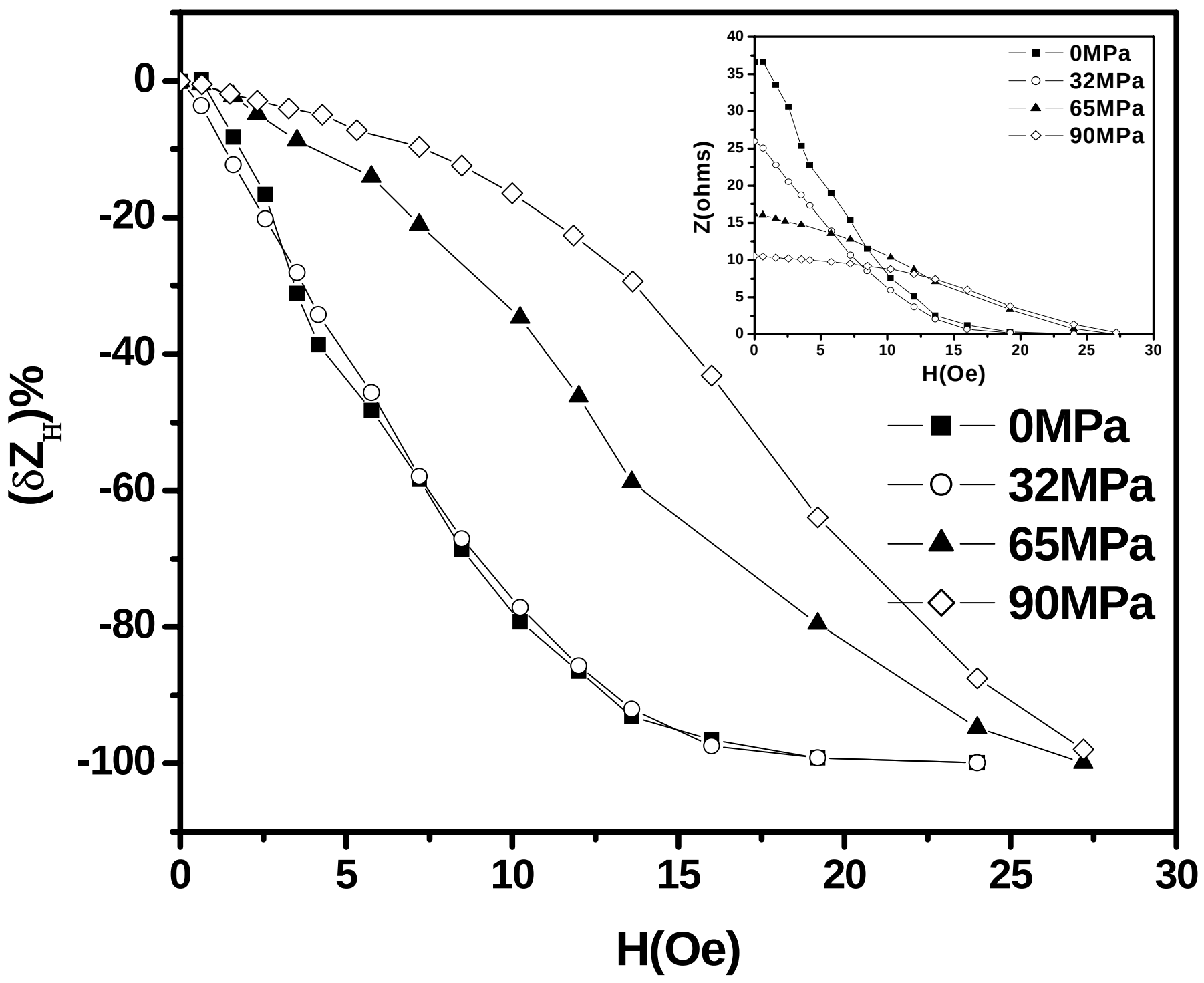

**Fig. 3. B. Kaviraj et al.**

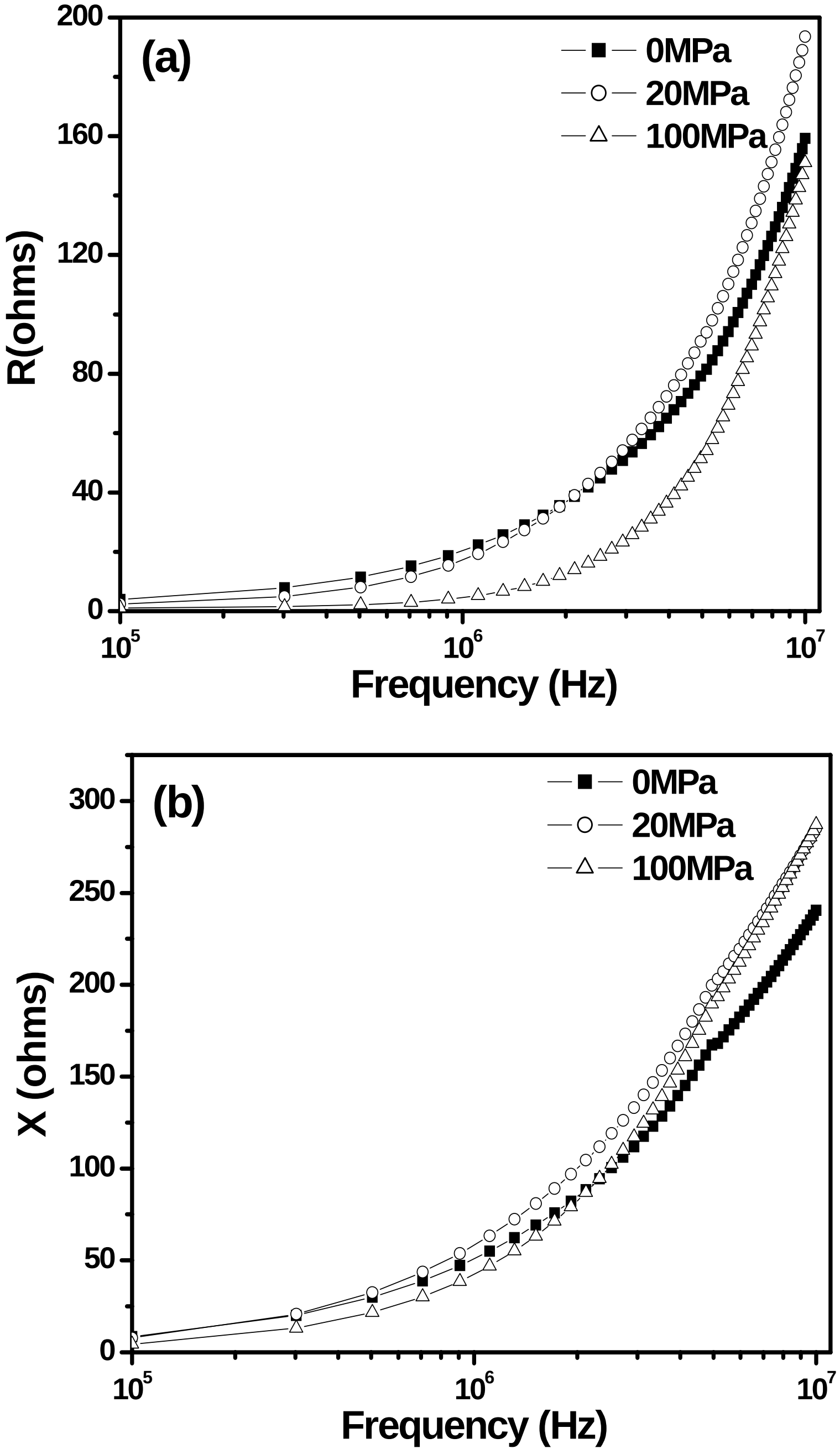

**Fig. 4. B. Kaviraj et al.**

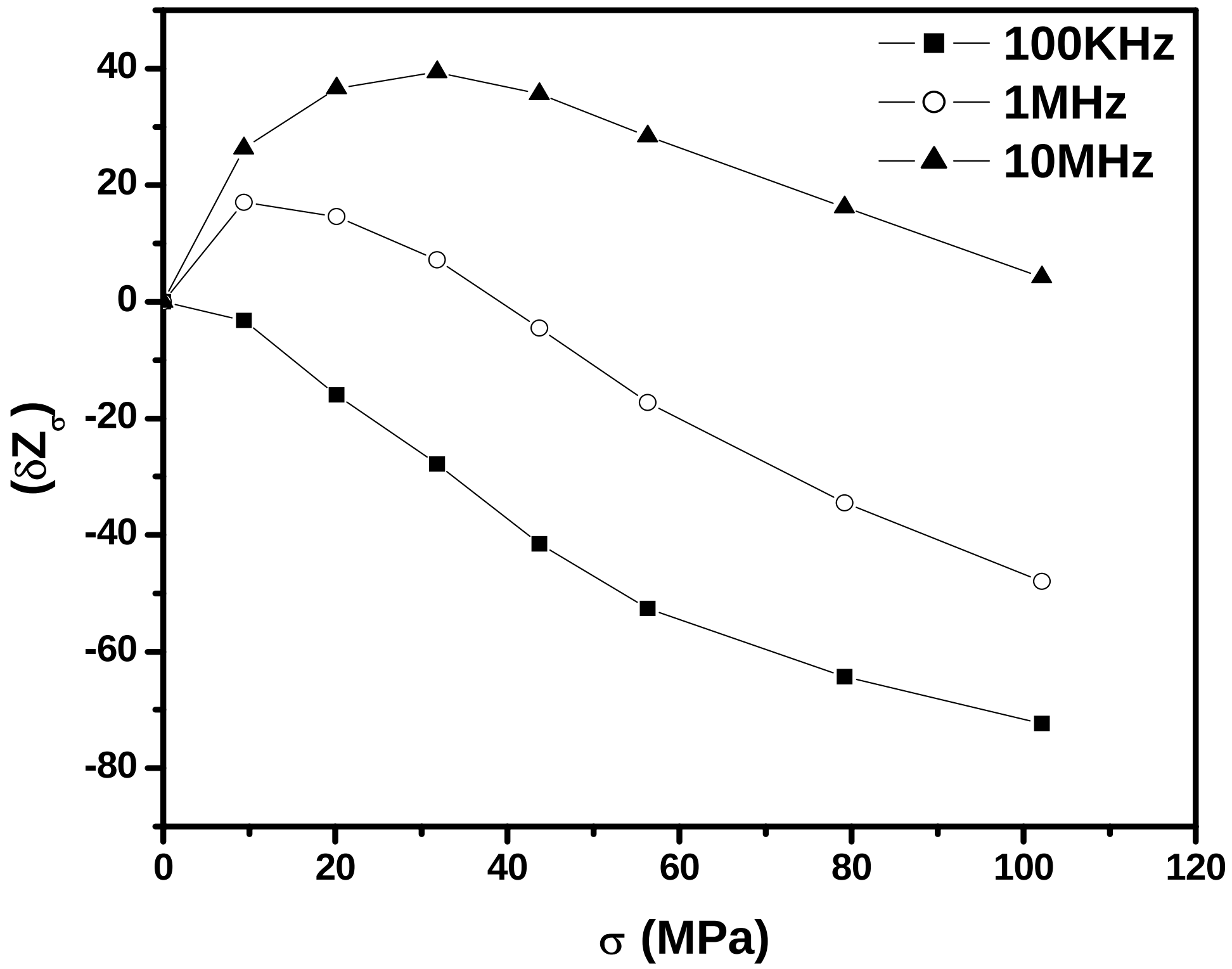

**Fig. 5. B. Kaviraj et al.**

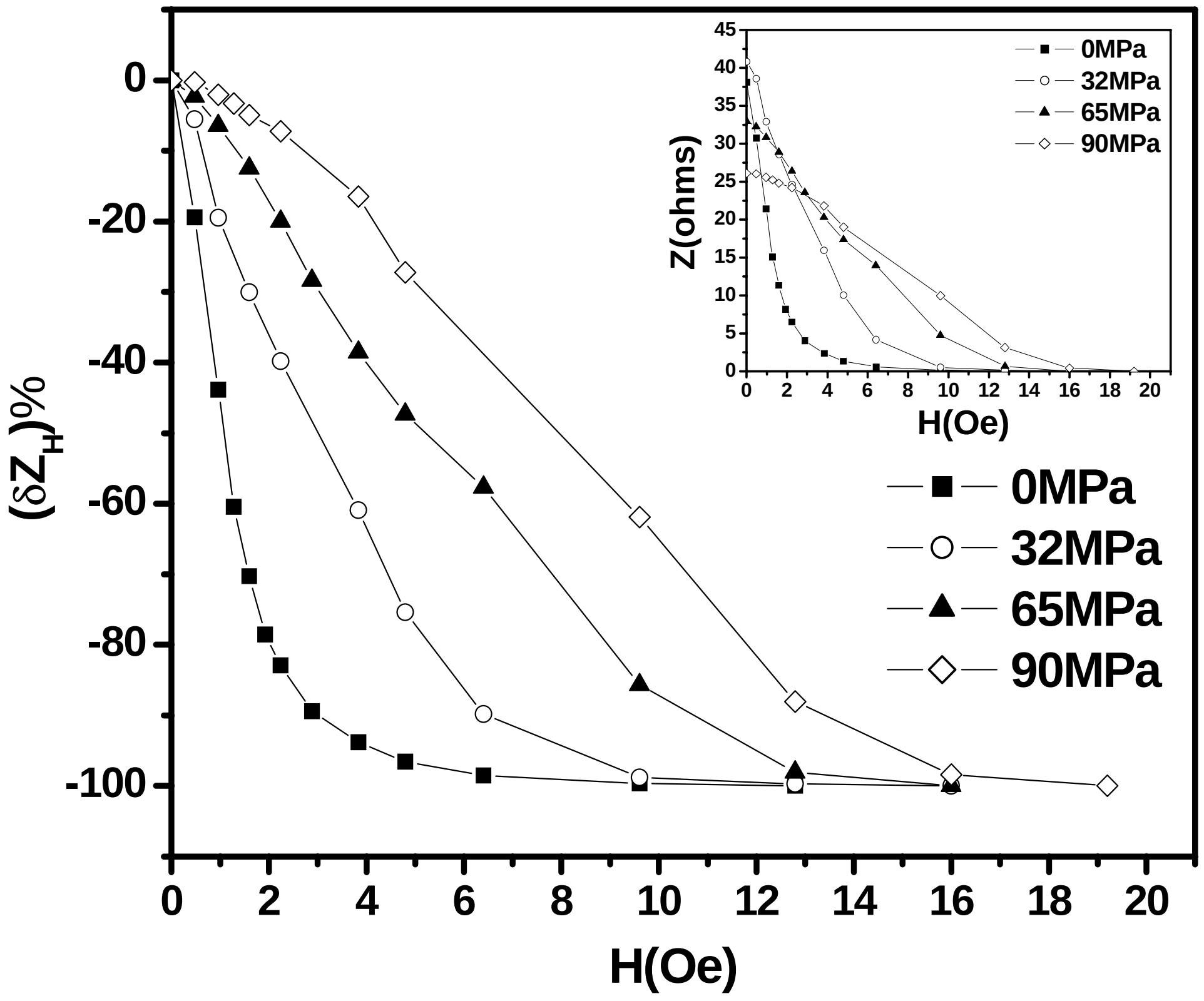

**Fig. 6. B. Kaviraj et al.**

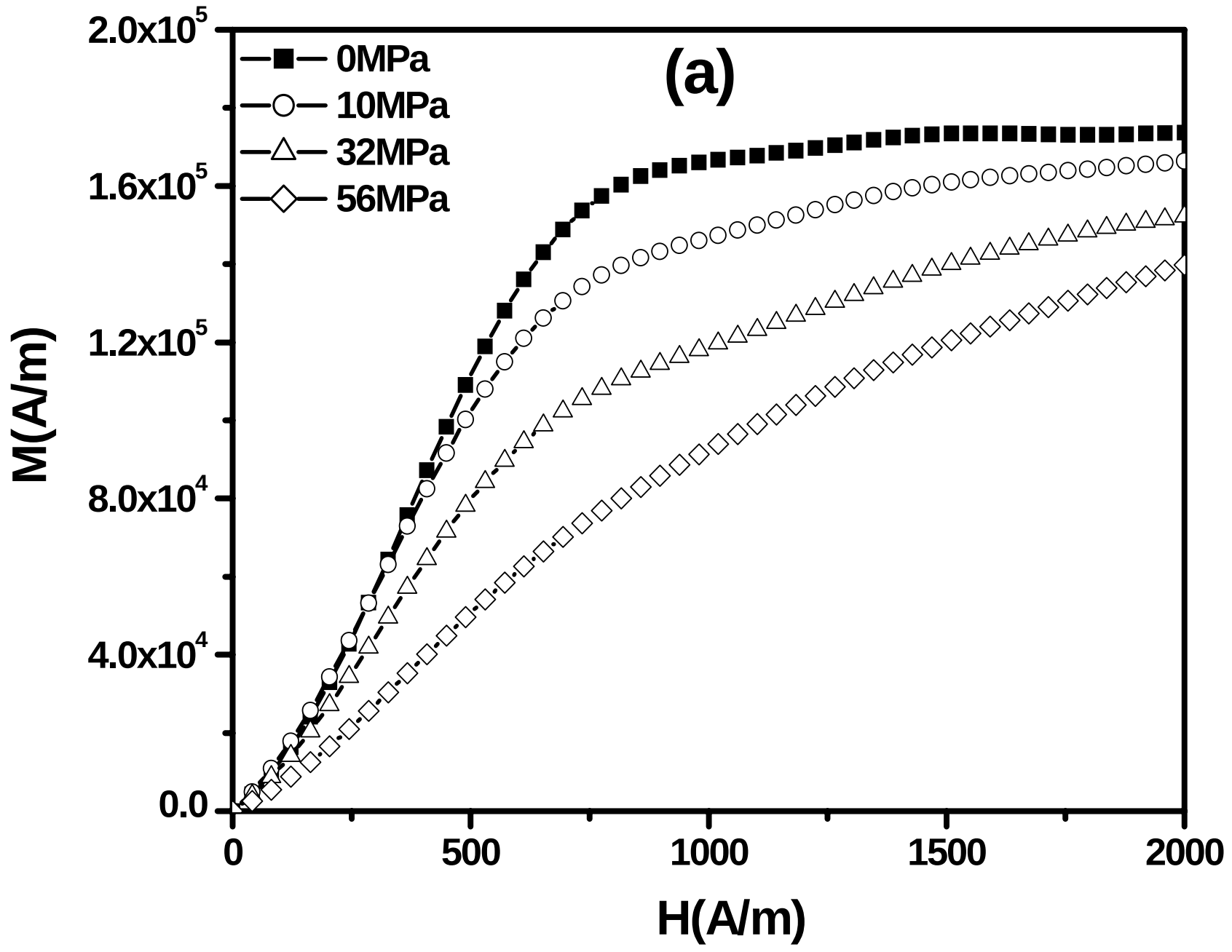

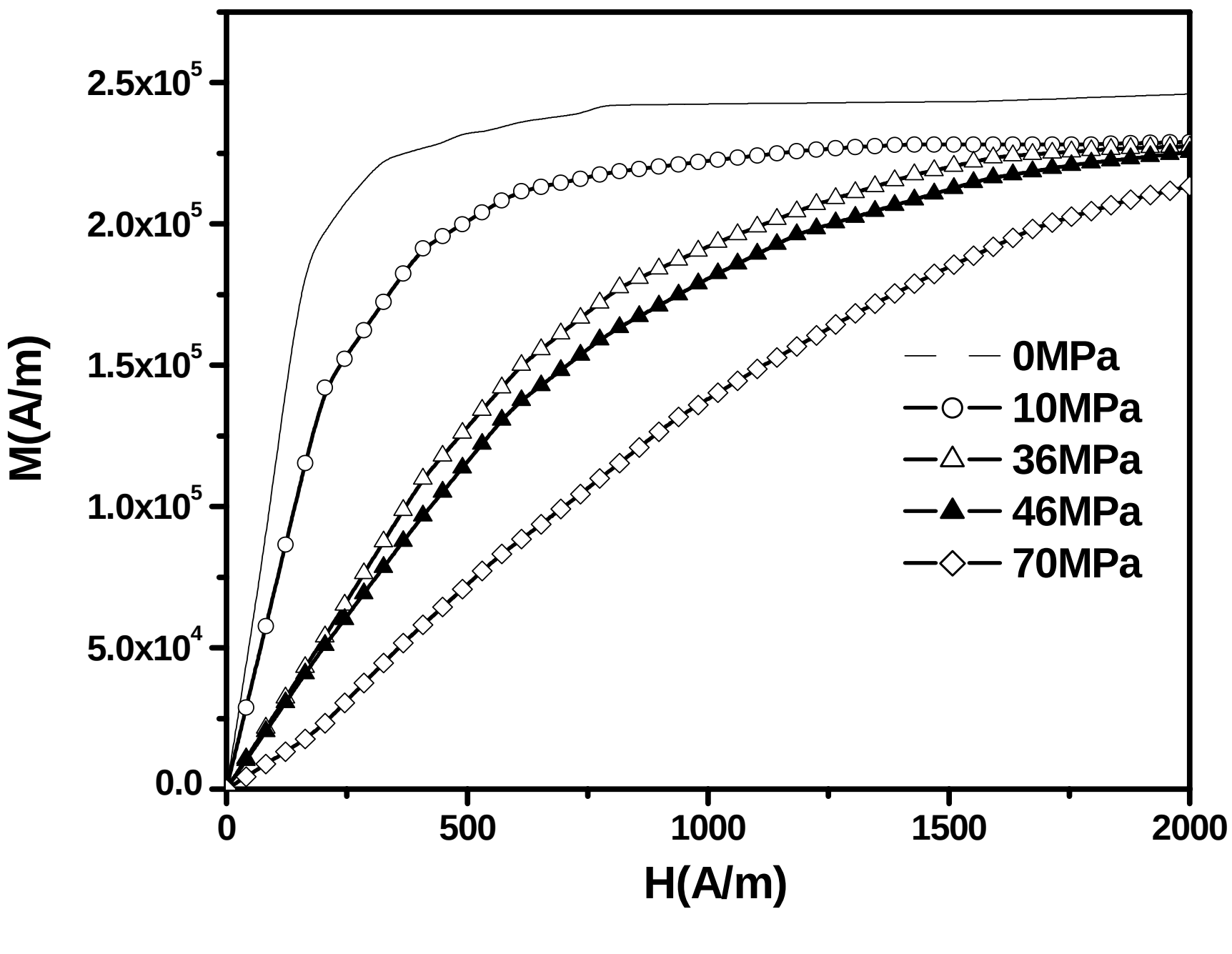

**Fig. 7. B. Kaviraj et al.**

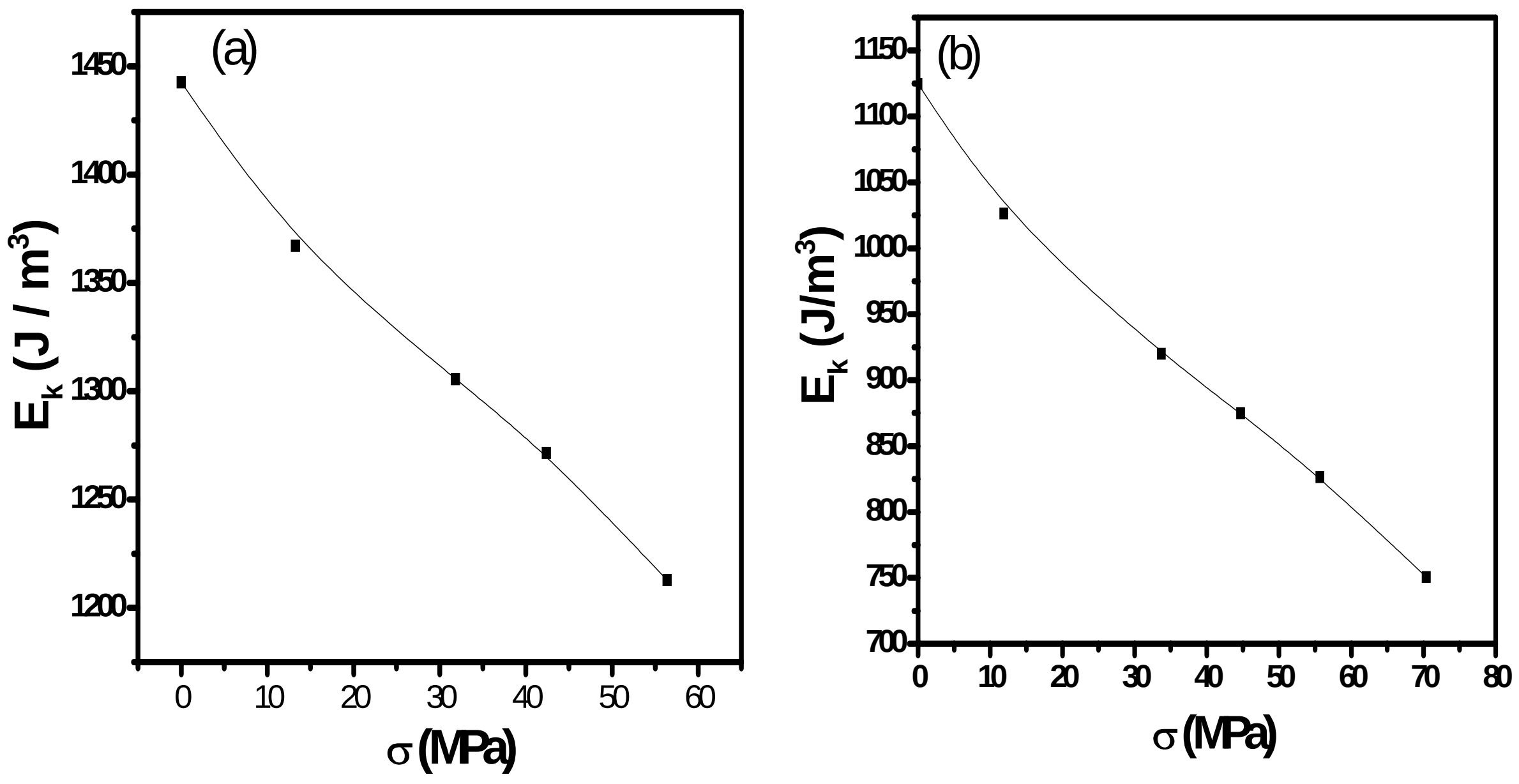

**Fig. 8. B. Kaviraj et al.**

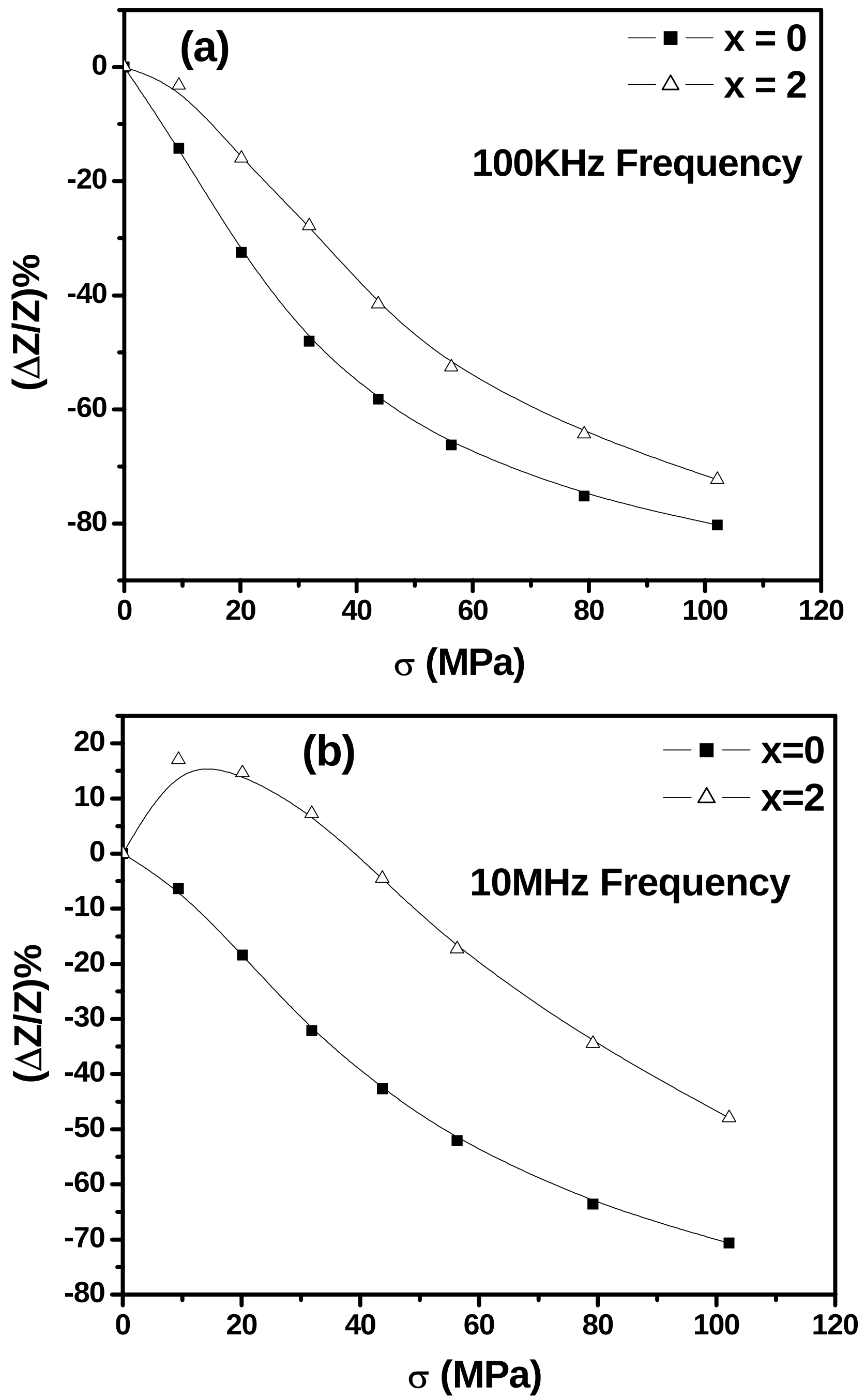

**Fig. 9. B. Kaviraj et al.**